\documentclass[aps,prb,twocolumn,showpacs,amsmath,amssymb]{revtex4}
\usepackage{txfonts}
\usepackage{graphicx}
\usepackage{cases}
\usepackage{dcolumn}
\usepackage{bm}

\begin{document}
\title{Interedge magnetic coupling in transition-metal terminated graphene nanoribbons}

\author{Yan Wang}
\author{Hai-Ping Cheng}
\affiliation{Department of Physics and Quantum Theory Project, University of Florida, Gainesville, Florida
32611, USA}

\begin{abstract}
The magnetic structures and interedge magnetic couplings of Fe, Co and Ni transition-metal terminated
graphene nanoribbons with zigzag (ZGNR) and armchair (AGNR) edges are studied by first-principles
calculations. Fe-ZGNR is found to show antiferromagnetic (AF) coupling between two edges, while the interedge
coupling of Co-ZGNR is ferromagnetic (FM). For Fe-AGNRs and Co-AGNRs, increasing the interedge distance we
follow oscillatory transitions from FM to AF coupling with a period of about 3.7 {\AA}. The damped
oscillatory behavior indicates a Ruderman-Kittel-Kasuya-Yosida type interedge magnetic coupling and the
oscillation period is determined by the critical spanning vector which connects two inequivalent Dirac points
in the graphene Brillouin zone. The two edges in Ni-ZGNR are decoupled independent of the ribbon width and
Ni-AGNRs are found to be nonmangetic.
\end{abstract}


\maketitle

Graphene, a monolayer of carbon atoms packed into a honeycomb lattice, continues to attract immense interest,
mostly because of its two-dimensional stability, unique band structure and other unusual physical
properties\cite{Geim:2009}. In particular, cutting graphene along two high-symmetry crystallographic
directions produces quasi-one-dimensional periodic strips of graphene with armchair or zigzag edges, usually
referred to as graphene nanoribbons (GNRs). The zigzag edge GNR (ZGNR) is theoretically predicted to be
magnetic with two spin-polarized edge states, which are ferromagnetically ordered but antiferromagnetically
coupled to each other through the graphene backbone \cite{Fujita:1996,Lee:2005,Son:2006}, while the armchair
edge GNR (AGNR) is found to be nonmagnetic. The interedge magnetic coupling in ZGNRs has attracted
considerable attention \cite{Lee:2005,Son:2006,Pisani:2007,Jung:2009}. Antiferromagnetic coupling of the two
zigzag edges in ZGNR can be explained in terms of interactions between the magnetic tails of the edge
states\cite{Lee:2005}, since the C atoms always stand at the opposite sublattices of GNR at the two zigzag
edges. The magnitude of the interedge magnetic coupling shows a $w^{-2}$ dependence as a function of the
ribbon width $w$ \cite{Pisani:2007,Jung:2009}.

Most of the studies of interedge magnetic coupling on GNRs focus on ribbons with zigzag edges and hydrogen
terminations. We have previously reported that metal terminated GNRs can also exhibit magnetics behavior
\cite{Wang:2010}. Furthermore, GNRs with ferromagnetically coupled edges terminated with transition metals
show high degree of spin polarization at the Fermi energy, and thus can be excellent candidate for spintronic
applications. The interedge magnetic coupling in Fe terminated ZGNRs has also been studied by Ong \emph{et
al.} \cite{Ong:2010} very recently, showing that the coupling is antiferromagnetic and the strength decreases
with increasing ribbon width.

For a full understanding of the interedge magnetic coupling in TM-GNR systems, in this paper, we present a
first-principles study of zigzag and armchair GNRs terminated with Fe, Co and Ni 3\emph{d} transition metals,
focusing on the width dependence of the magnetic coupling between two edges. Interestingly, the behavior of
interedge magnetic coupling is found to differ significantly with different type of metal terminations or
ribbon edges. We also find a damped oscillatory behavior of interedge magnetic coupling and the oscillation
period is determined by the critical spanning vector which connects two Dirac points in the graphene
Brillouin zone.

\begin{figure}
{\includegraphics[width=8.6cm]{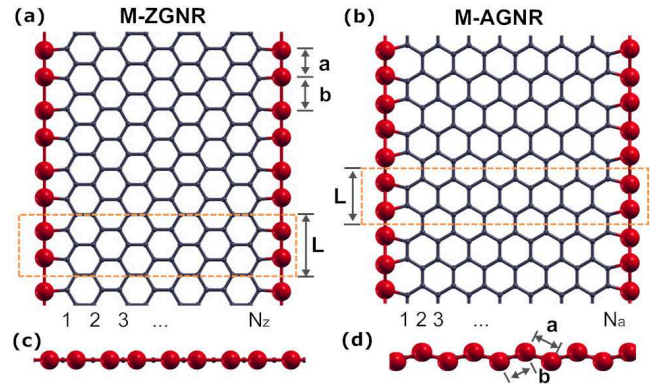}} \caption{\label{fig:geometry}(Color online). The top view
of structures of TM-ZGNR with $N_z=8$ (a) and TM-AGNR with $N_a=14$ (b). The solid (red) dots represent
transition-metal atoms passivating the ribbon edges. The dashed rectangle denotes the unit cell in
calculation. (c) and (d) are corresponding side views of the TM-ZGNR and TM-AGNR, respectively. $a$, $b$ and
$L$ are the two metal-metal bond lengths at each edge and the the length of unit cell, respectively.}
\end{figure}

\begin{table*}
\caption{\label{tab:table} Structural, energetic and magnetic properties of metal-terminated GNRs. Columns
show: metal termination, metal-metal ($a$ or $b$) and metal-carbon ($d_{\rm M-\rm C}$) bond length (in
{\AA}), the binding energy $E_b$ and the formation energy of the metal chain $E^{\rm chain}_{\rm formation}$
(in eV per metal atom), total energy difference between AFE and AF (in meV per unit cell) states, total
energy difference between FM and AF (in meV per unit cell) states, total magnetic moment $m_{\rm tot}$ of the
ribbon in the FM state (in Bohr magneton $\mu_{\rm B}$ per edge termination), local magnetic moment of the
metal atom $\mu_{\rm M}$ and its nearest-neighbor C atom $\mu_{\rm C}$ (in $\mu_{\rm B}$).}
\begin{ruledtabular}
\begin{tabular}{cccccccccccccc} & \textit{a} & \textit{b} & $d_{\rm M-\rm C}$ & $E_b$ & $E^{\rm chain}_{\rm formation}$ & $\Delta E^{\rm total}_{\rm AFE-AF}$ & $\Delta E^{\rm total}_{\rm FM-AF}$ & $m_{\rm
tot}$ & $\mu_{\rm M}$ & $\mu_{\rm C}$&\\
\hline
10-Fe-ZGNR& 2.18 & 2.76 & 1.86 & 4.10 & 1.63 & 1066.5 & 9.0 & 2.20 & 2.30 & $-0.14$\\
10-Co-ZGNR& 2.34 & 2.60 & 1.82 & 4.49 & 1.71 & 724.6 & -2.2 & 1.20 & 1.58 & $-0.06$ \\
10-Ni-ZGNR& 2.47 & 2.47 & 1.80 & 5.05 & 2.11 & $\backslash$ & $<$0.1 & 0.20 & 0.17 & $-0.01$ \\
17-Fe-AGNR& 2.14 & 2.41 & 1.86 & 3.71 & 1.84 & 911.6 & -4.6  & 2.12 & 2.14 & $-0.05$ \\
17-Co-AGNR& 2.23 & 2.35 & 1.83 & 3.99 & 2.07 & 363.5 & 4.6 & 1.19 & 1.54 & $-0.03$ \\
17-Ni-AGNR& 2.24 & 2.32 & 1.81 & 4.52 & 2.49 & $\backslash$ & $\backslash$ & 0 & 0 & 0\\
\end{tabular}
\end{ruledtabular}
\end{table*}

The electronic structure calculations are performed using density functional theory implemented in the
plane-wave-basis-set Vienna \textit{ab} initio simulation package (VASP) \cite{VASP:1996}. Each ribbon is
simulated within a supercell geometry containing 4 metal atoms at two edges, as shown in Fig.\
\ref{fig:geometry}. A large vacuum spacing of 15 {\AA} is used between two edges and between two graphene
planes to prevent interaction between adjacent images. Projector augmented wave (PAW) potentials with kinetic
energy cutoff of 500 eV are employed in all simulations. For the exchange and correlation functional we use
the Perdew-Burke-Ernzerh generalized gradient approximation \cite{Perdew:1996}. Brillouin-Zone sampling is
done on a grid of $36\times1\times1$ Monkhorst-Pack \cite{Monkhorst:1976} \emph{k}-points along the periodic
direction of the ribbon for ZGNRs and $40\times1\times1$ for AGNRs. The Gaussian smearing method is used to
treat partial occupancies, and the width of smearing is chosen to be 0.1 eV for geometry relaxations. The
geometries are optimized until all forces on all ions fall below the threshold value of 0.01 eV/{\AA}. To
obtain accurate magnetic configurations and ensure high accuracy in the calculated total energies of relaxed
structures, calculations are conducted using Gaussian smearing of 0.05 eV and PAW pseudopotentials for Fe and
Co which treat the 3\emph{p} semi-core states as valence states.

Two typical structures of metal-terminated GNRs with 8 carbon zigzag chains and 14 carbon dimer lines across
the ribbon width are shown in Fig.\ \ref{fig:geometry} (a) and (b), respectively. Hereafter we refer to a
metal-terminated GNR with $N_a$ dimer lines as a $N_a$-TM-AGNR and that with $N_z$ zigzag chains as a
$N_z$-TM-ZGNR where TM stands for Fe, Co or Ni. Both edges of the ribbon have the same configuration for each
case. We find the lowest-energy structures of the ribbon by structural relaxation calculations for three
possible termination configurations: the linear type ($a=b=L/2$), the dimerized linear type ($a\neq b$ and
$a+b=L$), and the zigzag type ($a+b>L$). The calculated ground state structural properties are listed in
Table \ref{tab:table} for 10-TM-ZGNRs and 17-TM-AGNRs. The results show that the most favorable structures
for all studied TM-ZGNRs are all linear while the TM-AGNRs are all zigzag type mainly because of a relatively
smaller length $L$ of the AGNR unit cell. The Fe and Co terminations dimerize at the edge of ribbon due to
the Peierls distortion \cite{Peierls:1955}. Changing the width of the ribbon will have negligible effect on
the structure of the edges. The binding energies $E_b$ of the metal atom, defined as $E_b=(E_{\rm
TM-ZGNR}-E_{\rm ZGNR})/4-E^{\rm atom}_{\rm M}$, and the formation energy of the metal chain, $E^{\rm
chain}_{\rm formation}=E^{\rm chain}_{\rm TM}/2-E^{\rm atom}_{\rm TM}$, are also listed in Table
\ref{tab:table}. The difference between $E_b$ and $E^{\rm chain}_{\rm formation}$ represents a direct binding
between the TM and carbon atoms at the ribbon edge. It is clearly shown that for all cases the metal atoms
bond strongly with edge carbon atoms.

Next we examine the magnetic structures and couplings between the magnetic moments in the ribbon edges. Three
states with different spin configurations of the TM terminations are considered: (i) antiferromagnetically
ordered spins at each edge of the ribbon, denoted by AFE, (ii) ferromagnetically ordered spins along both
edges with the same spin direction, denoted by FM, and (iii) ferromagnetically ordered spins at each edge
with the opposite spin directions between the edges, denoted by AF. Total energy calculations are performed
to decide the ground states of the magnetic structures.

Except for Ni-AGNR which is found to be nonmagnetic, each ribbon considered shows spin polarized edges with
ferromagnetic ordering of the metal atoms at each edge for the lowest-energy state. In Table \ref{tab:table}
we show examples of 10-TM-ZGNRs and 17-TM-AGNRs. Besides that the AFE state of Ni-ZGNR is found to be not
stable at all, it is clearly shown that AFE state with antiferromagnetically ordered spins at each edge are
not favored, with a large energy difference as compared to FM or AF states. Though in the Fe-AGNRs and
Co-AGNRs the metal terminations alternatively bond to the opposite sublattices of GNR along each armchair
edge, the favorable FM (or AF) state is consistent with the ferromagnetic ordering found in the ground states
for Fe, Co and Ni monatomic chains \cite{Ataca:2008}. In Fe-ZGNRs and Co-ZGNRs the energetic disadvantage of
AFE state is very obvious, and can be explained by the interactions between the magnetic tails of the edge
spins \cite{Lee:2005}, since the metal atoms always bond to the same sublattices of GNR at each zigzag edges.
Generally the magnetic moment of the ribbon comes mostly from the metal atoms and their nearest-neighbor C
atoms at the edges, and the edge C atom presents magnetization antiparallel to the nearby metal atom, as
shown in Table \ref{tab:table} for ribbons with FM states. In the AF state the moments at two edges have
exact the same values but with opposite signs, thus the net magnetic moment of the ribbon is zero.

\begin{figure}
{\includegraphics[width=8cm]{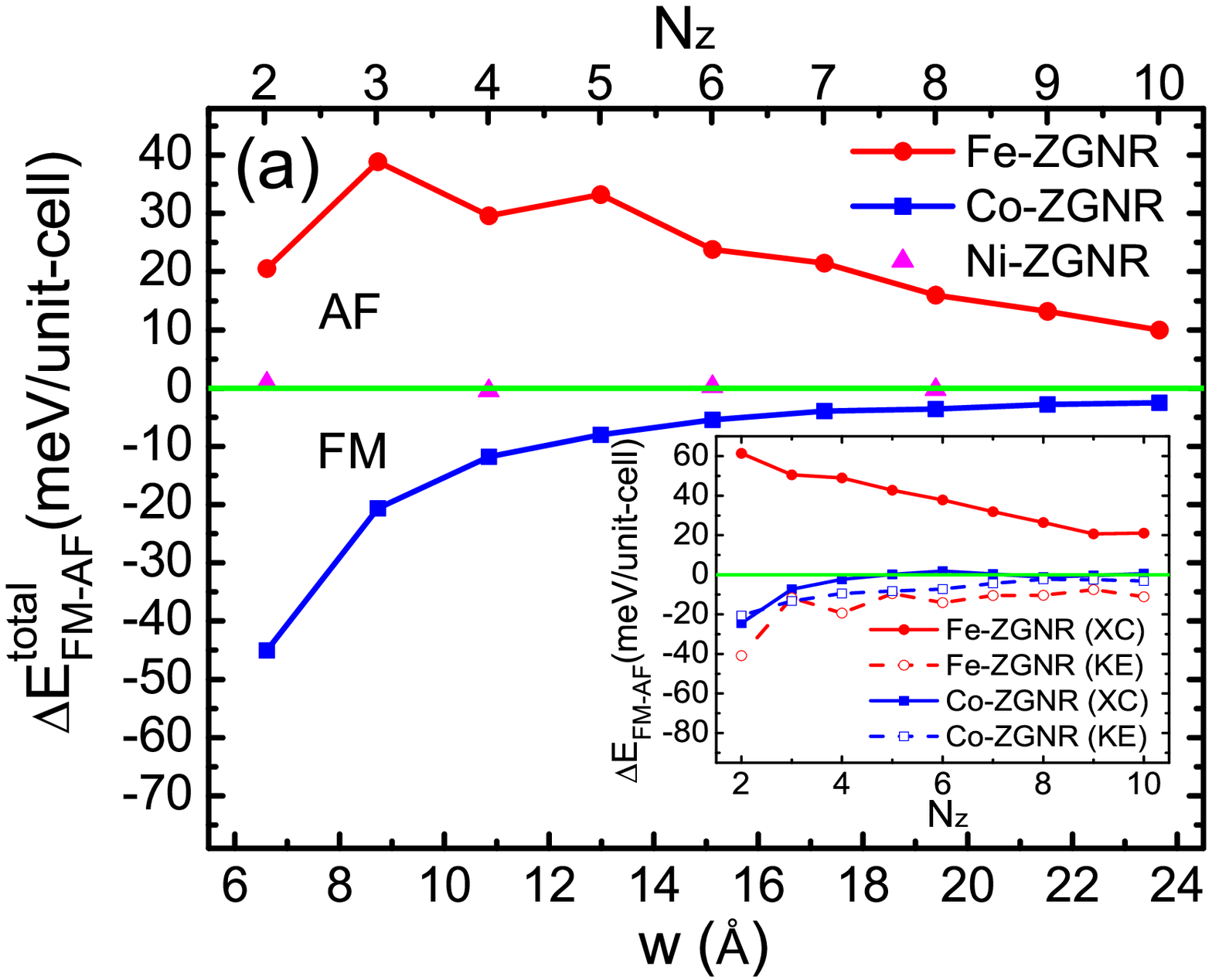}} {\includegraphics[width=8cm]{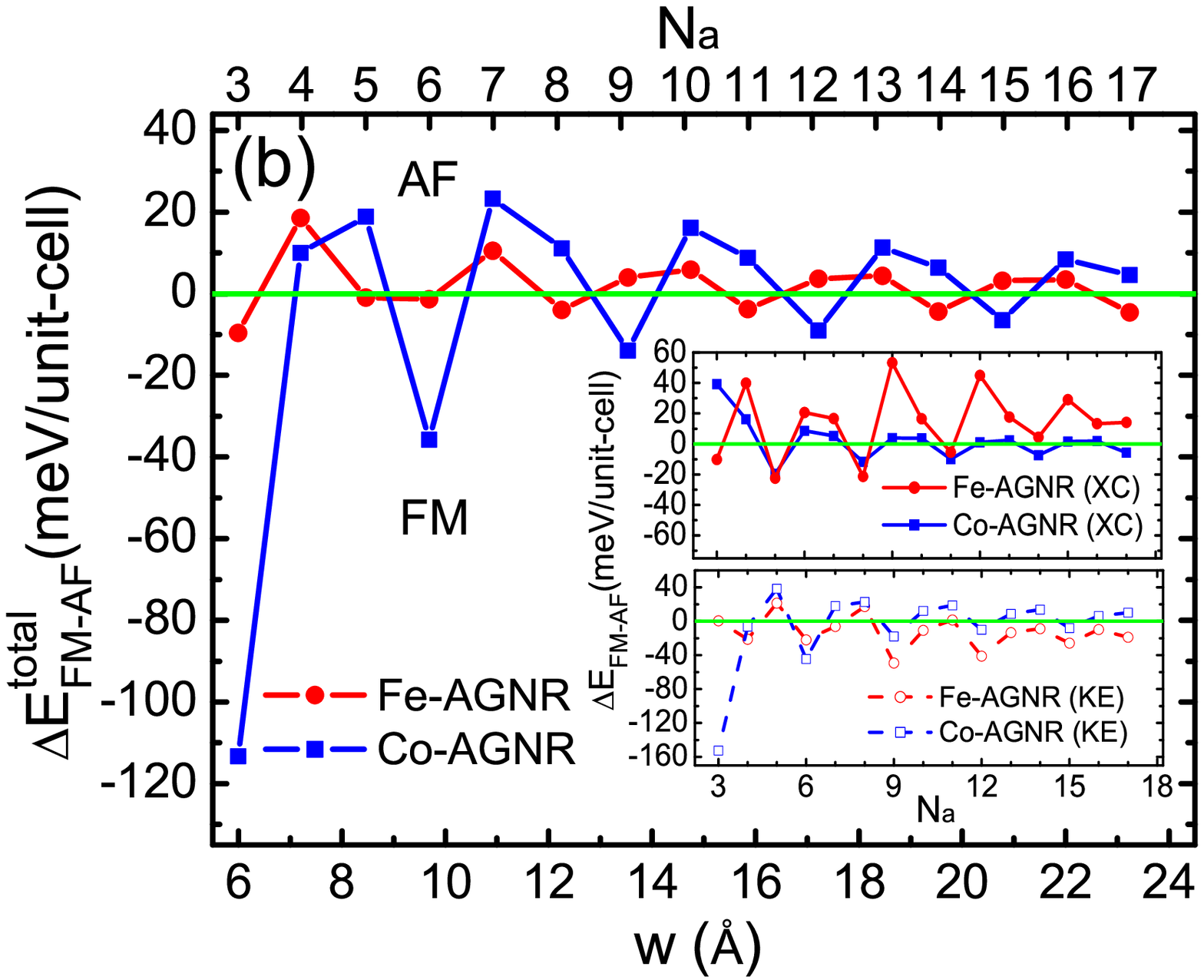}}
\caption{\label{fig:coupling}(Color online). Total energy difference between the FM and AF states as a
function of the ribbon width $w$. (a) Fe, Co and Ni terminated ZGNRs; (b) Fe and Co terminated AGNRs. Insets:
width dependence of the exchange-correlation (XC) contribution as well as kinetic and electrostatic (KE)
contribution to the total energy difference between the FM and AF states.}
\end{figure}

The magnetic order between the two edges for the ground state can be either FM or AF favored depending on the
interedge magnetic coupling. Flipping the spin moments of one of the edges will result a total energy change
in the system for comparable size of the magnetic moments at the metal terminations. The interedge magnetic
interaction strength can be identified by the total energy difference $\Delta E^{\rm total}_{\rm FM-AF}$
between the FM and AF states. Our calculations show that the ground states of Fe-ZGNRs are always AF, similar
to H-ZGNRs and ZGNRs without H-passivation \cite{Lee:2005,Son:2006,Pisani:2007}. The behavior of the $\Delta
E$ as a function of the ribbon width is shown in Fig.\ \ref{fig:coupling} (a). For $Nz > 4$ the $\Delta
E^{\rm total}_{\rm FM-AF}$ decreases almost linearly with increasing ribbon widths, in agreement with a
recent calculation by Ong \emph{et al.} \cite{Ong:2010}. However, in the case of 2-Fe-ZGNR the $\Delta E^{\rm
total}_{\rm FM-AF}$ is much lower than that of 3-Fe-ZGNR.

Contrary to the case of Fe-ZGNRs, we find that the interedge magnetic coupling of Co-ZGNR with a finite width
is always FM, as shown in Fig.\ \ref{fig:coupling} (a). The absolute value of difference in total energy also
decay as $N_z$ increases. It becomes negligible when the two edges are separated by a large ribbon width. By
fitting the $\Delta E^{\rm total}_{\rm FM-AF}$ variation with $w$, one obtains a decay law close to
$w^{-2.5}$. However, the almost monotonic width dependence of coupling for Fe-ZGNR and Co-ZGNRs shown in
Fig.\ \ref{fig:coupling} (a) are only part of the story and will be re-examined later. The smaller amplitude
of interedge magnetic interaction in Co-ZGNR at large $w$ as compared to Fe-ZGNR is primarily due to the fact
that magnetic moments of Co-ZGNRs are more localized localized at the edges \cite{Wang:2010}. Inside the
Co-ZGNR the magnetic tails of the edge moments decay much faster than Fe-ZGNR, resulting almost zero moments
in the inner C atoms of the ribbon. This localization effect is more remarkable in the case of Ni-ZGNRs. The
interedge exchange interaction is found to be negligible in Ni-ZGNRs, as the FM and AF states become almost
degenerate at even the shortest width with $Nz=2$.

The total energy difference between FM and AF states can be separated into non-interaction one-electron
(kinetic), Hartree (electrostatic) and exchange-correlation (XC) contributions:
\begin{equation}
\Delta E^{\rm total}_{\rm FM-AF}=\Delta E^{\rm Kinetic}_{\rm FM-AF}+\Delta E^{\rm Hartree}_{\rm FM-AF}+\Delta
E^{\rm XC}_{\rm FM-AF}.\label{eq:mynum}
\end{equation}

A close inspection of the $\Delta E^{\rm XC}_{\rm FM-AF}$ curves in Fig.\ref{fig:coupling} (a) shows that the
XC contribution to the total energy, is clearly responsible for the different type of interedge coupling in
Fe-ZGNRs and Co-ZGNRs. For Fe-ZGNRs the $\Delta E^{\rm XC}_{\rm FM-AF}$ is always positive, while in Co-ZGNRs
the $\Delta E^{\rm XC}_{\rm FM-AF}$ is negative but become negligible for large interedge distances. A
comparison between $\Delta E^{\rm XC}_{\rm FM-AF}$ and $\Delta E^{\rm total}_{\rm FM-AF}$ indicates that FM
coupling always lower the kinetic and electrostatic (KE) parts of the contribution for both Fe-ZGNRs and
Co-ZGNRs, especially for ribbons with short interedge distances. However, the two edges of Fe-ZGNR are still
AF coupled resulting from a large energy difference from the XC contribution. Take the 2-Fe-ZGNR as an
example: the FM coupling lowers the KE contributions to the total energy by 41 meV/unit-cell but at a cost of
$-61$ meV/unit-cell in the XC contribution, resulting a $\Delta E^{\rm XC}_{\rm FM-AF}=20$ meV/unit-cell for
an AF coupling as the ground state.

For GNRs with armchair edges terminated by Fe and Co atoms, the interedge coupling between the two edges
favor either AFM or FM depending on the ribbon width. The energy difference between the FM and AF states as a
function of the ribbon width is plotted in Fig.\ref{fig:coupling} (b). A damped oscillatory behavior, as
shown in Fig.\ref{fig:coupling} (b) for both Fe-AGNRs and Co-AGNRs, clearly indicate a
Ruderman-Kittel-Kasuya-Yosida (RKKY)-type \cite{RKKY} interedge exchange. A $N_a$-Fe-AGNR is FM coupled only
if $N_a= 3m+2$ (where m is a positive integer), and a $N_a$-Co-AGNR is FM only if $N_a= 3m$.

\begin{figure}
{\includegraphics[width=7.8cm]{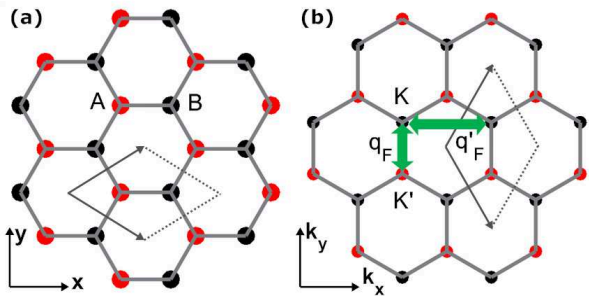}} \caption{\label{fig:graphene}(Color online). (a) Real
space structure of the honeycomb lattice with two sublattices A and B in red (light) and black (dark) colors
respectively. (b) The \emph{critical spanning vectors} $q_{\rm F}$ and $q{'}_{\rm F}$ for graphene in
reciprocal space. The $q_{\rm F}$ connects two inequivalent Dirac point $K$ and $K{'}$ and parallel to the
$k_y$ axis which is the metal-AGNR interface (along $x$ axis in real space) normal. The $q{'}_{\rm F}$
connects two equivalent Dirac point $K$ and $K$ (or $K{'}$ and $K{'}$) and parallel to the $k_x$ axis which
is the metal-ZGNR interface (along $y$ axis in real space) normal.}
\end{figure}

The well-known long-range RKKY interaction between two magnetic impurities in a non-magnetic host material is
mediated by the conduction electrons of the host, and the coupling strength $J$ can be written as
\cite{Fischer:1975,Yafet:1987}
\begin{equation}
J(w)=J_0 \cos(2 k_{\rm F} w+\phi)/r^D,
\end{equation}
where $D$ is the assumed dimensionality, $w$ is the distance between two impurities and $k_{\rm F}$ is the
wavevector at the Fermi level. By choosing dimensionality $D=1$ and set $w$ as the interedge distance of the
ribbon, our results of the total energy difference $\Delta E^{\rm total}_{\rm FM-AF}(w)\propto J(w)$ can be
fitted very well using Eq. (2) with a period $T_{\rm A}=\pi/k_{\rm F}$ of about 3.7 {\AA} for both Fe-AGNRs
and Co-AGNRs and slightly different phase $\phi$. This is analogous to the case of the interlayer exchange
coupling between ferromagnetic layers separated by nonmangnetic metallic spacer \cite{Stiles:1999}, where a
two-dimensional range function, $J(d)=J_0 \cos(q_{\rm F} d+\phi)/d^2$, is used to describe the oscillatory
behavior of the coupling strength as a function of the spacer thickness $d$. The $q_{\rm F}$, which
determines the oscillation period, is the \emph{critical spanning vector} parallel to the interface normal
that connects two sheets of the Fermi surface of the spacer at a point where they are parallel to each other.
Similarly, here we find that the fitted $2k_{\rm F}=q_{\rm F}$ is exactly the \emph{critical spanning vector}
connecting two inequivalent Dirac points $K$ and $K{'}$ in the Brillouin zone of graphene in the direction of
the metal-AGNR interface (along the armchair edges) normal, which is shown in Fig.\ref{fig:graphene}. Another
spanning vector $q{'}_{\rm F}$, which connects two equivalent Dirac points $K$-$K$ or $K{'}$-$K{'}$,
determines the oscillation period $T_{\rm Z}$ of the interedge coupling for ZGNRs. However, following the
discussion above we get $T_{\rm Z}=2\pi/q{'}_{\rm F}=2.1$ {\AA} which coincides with the interedge lattice
spacing in ZGNRs. This clearly explains why the damped oscillatory behavior is not shown in actual width
dependence of the $\Delta E^{\rm total}_{\rm FM-AF}$ for either Fe-ZGNRs or Co-ZGNRs but a monotonic
behavior.

The width dependences of the XC and KE contributions to the total energy difference are also shown in
Fig.\ref{fig:coupling} (b) for Fe-AGNRs and Co-AGNRs. Both $\Delta E^{\rm XC}_{\rm FM-AF}$ and $\Delta E^{\rm
XC}_{\rm FM-AF}$ show similar oscillatory behavior as the $\Delta E^{\rm total}_{\rm FM-AF}$ with a same
period, a phase shift as compared to the $\Delta E^{\rm total}_{\rm FM-AF}$ curve is found for the KE
contribution in Fe-AGNRs, and for the XC contribution in Co-AGNRs. This indicates that the damped oscillatory
behavior in the width dependence of $\Delta E^{\rm total}_{FM-AF}$ is resulting from a competition between
different contributions to the total energy.

In summary, we have presented a study of the interedge magnetic coupling of Fe, Co and Ni terminated graphene
nanoribbons through first-principles calculations. We find that the ferromagnetic ordering of the metal
terminations at each edge of the ribbon is favored for both ZGNRs and AGNRs. Whether the interedge magnetic
coupling is ferromagnetic or antiferromagnetic depends to a large extent on the type of metal atoms and
edges, as well as the ribbon width. The two edges for Fe-ZGNR are found to be antiferromagnetically coupled
while for Co-ZGNR ferromagnetic coupling is favored, and the strength of the interedge exchange interaction
decreases as the ribbon width increases. For both Fe-AGNRs and Co-AGNRs the interedge exchange interactions
show damped oscillatory behavior as a function of the ribbon width with a period of about 3.7 {\AA}. The
interedge magnetic coupling is negligible in Ni-ZGNRs, and Ni-AGNRs are found to be nonmagnetic.

This work was supported by US/DOE/BES/DE-FG02-02ER45995. The authors acknowledge DOE/NERSC and the UF-HPC
center and for providing computational resources.

\end{document}